\documentclass[aps,showpacs,preprintnumbers, superscriptaddress, nofootinbibt,twocolumn]{revtex4}
\usepackage{amsmath}
\usepackage{eurosym}
\usepackage{amssymb}

\pdfoutput=1

\usepackage{graphicx}

\usepackage{color}

\usepackage{braket}

\def\be{\begin{equation}}
\def\ee{\end{equation}}
\def\bea{\begin{eqnarray}}
\def\eea{\end{eqnarray}}

\begin{document}

\title{Is there a connection between ``dark" and ``light" physics?}
\author{Matthew J. Lake}
\email{matthewj@nu.ac.th}
\affiliation{The Institute for Fundamental Study, ``The Tah Poe Academia Institute", Naresuan University, Phitsanulok 65000, Thailand}
\affiliation{Thailand Center of Excellence in Physics (ThEP), Ministry of Education, Bangkok 10400, Thailand}
\date{\today}

\begin{abstract}
In the early-mid 20$^{\rm th}$ century Dirac and Zel'dovich were among the first scientists to suggest an intimate connection between cosmology and atomic physics. Though a revolutionary proposal for its time, Dirac's Large Number Hypothesis (1937) adopted a standard assumption of the day, namely, the non-existence of the cosmological constant term ($\Lambda = 0$). As a result, its implementation necessitated extreme violence to the theory of  general relativity -- something few physicists were prepared to sacrifice in favour of `numerology' -- requiring a time-dependent gravitational coupling of the form $G(t) \sim 1/t$. Zel'dovich's insight (1968) was to realise that a small but nonzero cosmological term ($\Lambda > 0$) allowed the present day radius of the Universe to be identified with the de Sitter radius, $r_{\rm U} \simeq l_{\rm dS} \simeq 1/{\sqrt{\Lambda}}$, which removed the need for time-dependence in the fundamental couplings. Thus, he obtained the formula $\Lambda \simeq m^6G^2/\hbar^4$, where $m$ is a mass scale characterising the relative strengths of the gravitational and electromagnetic interactions, which he identified with the proton mass $m_{\rm p}$. In this paper, we review a number of recent arguments which, instead, suggest the identification $m = m_{\rm e}/\alpha_{\rm e}$, where $m_{\rm e}$ is the electron mass and $\alpha_{\rm e} = e^2/\hbar c \simeq 1/137$ is the usual fine structure constant. We note that these are of a {\it physical} nature and, therefore, represent an attempt to lift previous arguments \emph{{\` a} la} Dirac from the realm of numerology into the realm of empirical science. If valid, such arguments suggest an intimate connection, not only between the macroscopic and microscopic worlds, but, perhaps even more surprisingly, between the very essence of ``dark" and ``light" physics. 

{\textbf{Keywords}: Large Number Hypothesis, Dark Energy, Minimum Length Uncertainty Relations, Generalized Uncertainty Principle, Quantum Gravity, Neutrino Mass Problem, Holography, Variation of Fundamental Constants}
\end{abstract}

\pacs{01.30.Cc; 01.65.+g; 04.20.Cv; 04.50.Kd; 04.60.Bc; 05.10.Cc; 89.70.+c; 95.36.+x}
\maketitle

%Table of Contents%%%%%%%%%%%%%%%%%%%%%%%%%%%%%%%%%%%%%%%%%%
%%%%%%%%%%%%%%%%%%%%%%%%%%%%%%%%%%%%%%%%%%%%%%%%%%%%
\tableofcontents

%Sec.1%%%%%%%%%%%%%%%%%%%%%%%%%%%%%%%%%%%%%%%%%%%%%%%%
%%%%%%%%%%%%%%%%%%%%%%%%%%%%%%%%%%%%%%%%%%%%%%%%%%%%
\section{Introduction -- Dirac, Zel'dovich and the Large Number Hypothesis} \label{sect1}

In 1937, Dirac noted the approximate order of magnitude equivalence between several large dimensionless numbers obtained from atomic physics and cosmology \cite{Dirac1937}. These included the ratio of the present day radius of the Universe, $r_{\rm U} \simeq 4.40 \times 10^{28}{\rm cm}$, to the classical electron radius, $r_{\rm e} = 2.818 \times 10^{-13}{\rm cm}$, and the ratio of the electric and gravitational forces between an electron and a proton, 
\begin{eqnarray}  \label{LNH-1}
\frac{r_{\rm U}}{r_{\rm e}} \simeq 10^{40} \cong \frac{e^2}{Gm_{\rm e}m_{\rm p}} \simeq 10^{39} \, ,
\end{eqnarray}
where $m_{\rm e} = 9.109 \times 10^{-28}{\rm g}$ and $m_{\rm p} = 1.672 \times 10^{-24}{\rm g}$. (For convenience, from here on, we use the most recent measured, or inferred, values of physical length and mass scales.) Interpreting this as a signature of an as yet unknown connection between cosmological and atomic physics, he formulated the Large Number Hypothesis (LNH), which states that the numerical equality between two very large quantities with similar physical meanings cannot be a simple coincidence \cite{Dirac1937,Dirac1938,Dirac1974,Dirac1979}. However, since the radius of the Universe is not constant but scales approximately as $r_{\rm U}(t) \sim t$, Dirac surmised that the gravitational coupling varies according to $G(t) \sim 1/t$ -- under the assumption $\Lambda = 0$ -- though this is, of course, not the only way to satisfy Eq. (\ref{LNH-1}) for all epochs \cite{Ray:2007cc}. Nonetheless, such variation is strongly at odds with the current `concordance' model of cosmology, which is based on Einstein's theory with a positive cosmological term and an unknown ``dark" matter component \cite{Ostriker:1995rn}.

In 1968, ZelÕdovich noted the same approximate equivalence between the ratio of the radius of the Universe and the Compton wavelength of the proton, $\lambda_{\rm p} = h/m_{\rm p}c = 1.321 \times 10^{-13}{\rm cm}$, and between $\lambda_{\rm p}$ and the proton's Schwarzschild radius, $r_{\rm S}(m_{\rm p}) = 2Gm_{\rm p}/c^2 = 2.484 \times 10^{-52}{\rm cm}$ \cite{Zel'dovich:1968zz}. In addition, he noted that, if $\Lambda > 0$ and $r_{\rm U} \simeq l_{\rm dS} \simeq 1/\sqrt{\Lambda}$ -- contrary to Dirac's original assumptions -- then
\begin{eqnarray}  \label{LNH-2}
\frac{r_{\rm U}}{\lambda_{\rm p}} \simeq \frac{m_{\rm p} c}{h \sqrt{\Lambda}}\simeq 10^{40}  \cong \frac{\lambda_{\rm p}}{r_{\rm S}(m_{\rm p})} \simeq \frac{h c}{G m_{\rm p}^2} \simeq 10^{39} \, , 
\end{eqnarray}
and hence
\begin{eqnarray}  \label{LNH-3}
\Lambda \simeq \frac{m_{\rm p}^6G^2}{h^4} \simeq 10^{-56} {\rm cm^{-2}} \, . 
\end{eqnarray}
Setting $\Lambda \simeq1/l_{\rm dS}^2$, where $l_{\rm dS}$ is the de Sitter horizon, this is equivalent to
\begin{eqnarray} \label{LNH-4}
\lambda_{\rm p} = \frac{h}{m_{\rm p}c} \simeq (r_{\rm Pl}^2l_{\rm dS})^{1/3} \, , 
\end{eqnarray}
with $l_{\rm dS} \simeq 10^{28}{\rm cm}$, where $r_{\rm Pl} \equiv \sqrt{hG/c^3}$ is the Planck length. Zel'dovich's formula (\ref{LNH-3}) is remarkable in that, not only does it establish a possible connection between dark energy ($\Lambda$), canonical gravity ($G$) and elementary particle physics ($m_{\rm p})$, it is {\it also} compatible with current experimental bounds on the dark energy density \cite{Betoule:2014frx,PlanckCollaboration}.

The current best fit to the available cosmological data favours a cosmological concordance, or $\Lambda$CDM model, in which dark energy takes the form of a positive cosmological constant, which accounts for approximately 69$\%$ of the total energy density of the Universe, whereas cold dark matter (CDM) accounts for around 26$\%$ and ordinary visible matter for around 5$\%$ \cite{Betoule:2014frx,PlanckCollaboration}. The present day density is close to the critical density, 
\begin{eqnarray}  \label{rho_crit}
\rho_{\rm crit} = \frac{3{\cal H}_0^2}{8\pi G} = 8.639 \times 10^{-30} \rm g cm^{-3} \, ,
\end{eqnarray}
where ${\cal H}_0 = 67.74 \rm kms^{-1}Mpc^{-1} = 2.198 \times 10^{-18} \ \rm s^{-1}$ is the Hubble constant, so that the dark energy density is
\begin{eqnarray} \label{rho_Lambda}
\rho_{\Lambda} = \frac{\Lambda c^2}{8\pi G} = 5.971 \times 10^{-30} {\rm g cm^{-3}} \, .
\end{eqnarray}
This yields an inferred value of the cosmological constant,
\begin{eqnarray} \label{Lambda_est}
\Lambda = 1.114 \times 10^{-56} \rm cm^{-2} \, ,
\end{eqnarray}
which corresponds to a de Sitter radius, $l_{\rm dS} = 1.641 \times 10^{28}{\rm cm}$.

We now note the approximate equivalence between the classical electron radius and the proton Compton wavelength, 
\begin{eqnarray} \label{}
&&\lambda_{\rm p} = \frac{h}{m_{\rm p}c^2} = 1.321 \times 10^{-13}{\rm cm} 
\nonumber\\
&\simeq& r_{\rm e} = \frac{e^2}{m_{\rm e}c^2} = 2.818 \times 10^{-13}{\rm cm} \, , 
\end{eqnarray}
or $\lambda_{\rm p} \simeq r_{\rm e} = \alpha_{\rm e}k^{-1}_{\rm e}$, where $k^{-1}_{\rm e} = 2\pi/\lambda_{\rm e} = \hbar/m_{\rm e}c$ is the {\it reduced} Compton wavelength of the electron and $\alpha_{\rm e} = e^2/\hbar c = 7.297 \times 10^{-3}$ is the fine structure constant. Performing the replacement $\lambda_{\rm p} \rightarrow r_{\rm e}$ in Eq. (\ref{LNH-3}) then yields
\begin{eqnarray}  \label{Lambda}
\Lambda \simeq 
%\frac{l_{\rm Pl}^4}{r_{\rm e}^6} = 
\frac{m_{e}^6G^2}{\alpha_{e}^6\hbar^4} \simeq 1.366 \times 10^{-56} {\rm cm^{-2}} \, ,
\end{eqnarray}
or, equivalently,
\begin{eqnarray} \label{r_e}
r_{\rm e} = \frac{e^2}{m_{\rm e}c^2} \simeq (l_{\rm Pl}^2l_{\rm dS})^{1/3} \, , 
\end{eqnarray}
where $l_{\rm Pl} = \sqrt{\hbar G/c^3}$ denotes the {\it reduced} Planck length. 

Remarkably, since the publication of Zel'dovich's seminal paper \cite{Zel'dovich:1968zz}, the relations (\ref{Lambda})-(\ref{r_e}) have been obtained in the literature using at least {\it four} independent, yet not inconsistent, methods. In the present work, we review these four main, existing approaches, and briefly discuss directions for future research. The outline of this paper is as follows. In Sec. \ref{sect2}, the four approaches are reviewed in chronological order (i.e., the order in which they were proposed in the literature), in subsections \ref{sect2.1}-\ref{sect2.4}, respectively. In Sec. \ref{sect3}, the implications of the relations (\ref{Lambda})-(\ref{r_e}) for contemporary issues in theoretical physics, including holography \cite{tHooft:1999rgb,Bousso:2002ju} and the present day accelerated expansion of the Universe \cite{Betoule:2014frx,PlanckCollaboration}, are considered, along with potentially novel implications for the early Universe. A brief summary of our conclusions and main results is given in Sec. \ref{sect4}.

%Sec.2%%%%%%%%%%%%%%%%%%%%%%%%%%%%%%%%%%%%%%%%%%%%%%%%
%%%%%%%%%%%%%%%%%%%%%%%%%%%%%%%%%%%%%%%%%%%%%%%%%%%%
\section{Beyond numerology -- physical arguments for the LNH} \label{sect2}

%Sec.2.1%%%%%%%%%%%%%%%%%%%%%%%%%%%%%%%%%%%%%%%%%%%%%%%%
\subsection{Renormalization group flow of the vacuum energy} \label{sect2.1}

The relations (\ref{Lambda})-(\ref{r_e}) were first obtained in 1993 by Nottale \cite{Nottale} who argued that, like other fundamental `constants', the cosmological constant is an explicitly scale-dependent quantity, obeying a renormalization group equation. As such, its present day value may be split into a `bare' gravitational part and a scale-dependent, quantum mechanical vacuum energy part, i.e. $\Lambda(r) = \Lambda_{\rm G} + \Lambda_{\rm Q}(r)$. 

Following Zel'dovich \cite{Zel'dovich:1968zz}, who also noted that the bare zero-point energy is unobservable, and who suggested that the {\it observable} contribution to the vacuum energy is given by the gravitational energy of virtual particle-antiparticle pairs, continually created and annihilated in the vacuum state, 
\bea \label{N-1}
E_{\rm grav} \simeq \frac{Gm^2(r)}{r} \, , 
\eea
where $m(r) \simeq \hbar/(cr)$ is the effective mass of the particles at scale $r$, Nottale obtained the scale-dependent formula for the vacuum energy density as
\bea \label{N-2}
\rho_{\rm vac}(r) \simeq \rho_{\rm Pl}\left(\frac{l_{\rm Pl}}{r}\right)^6 \, . 
\eea
Here, $\rho_{\rm Pl} = (3/4\pi)m_{\rm Pl}/l_{\rm Pl}^3$ is the Planck density, and $m_{\rm Pl} = \sqrt{\hbar c/G}$ denotes the reduced Planck mass. 

Further assuming a renormalization group equation of the form
\bea \label{N-3}
\frac{d\rho_{\rm vac}}{d\ln(r)} = \gamma(\rho_{\rm vac}) \, , 
\eea
where $\gamma(\rho_{\rm vac})$ is an {\it unknown} function, which may be expanded as $\gamma(\rho_{\rm vac}) \simeq \gamma_0 + \gamma_1 \rho_{\rm vac}$ to first order, for $\rho_{\rm vac} \lesssim \rho_{\rm Pl}$, yields
\bea \label{N-4}
\rho_{\rm vac}(r) \simeq \rho_0\left[1 + \left(\frac{r_0}{r}\right)^{-\gamma_1}\right] \, , 
\eea
where $\rho_0 = -\gamma_1/\gamma_0$ and $r_0$ is an integration constant. Comparison of Eqs. (\ref{N-2}) and (\ref{N-4}) then gives $\gamma_1 = -6$, $r_0=l_{\rm Pl}$ and $\rho_0 = \rho_{\rm Pl}$ ($\gamma_0 = 6/\rho_{\rm Pl}$). 

Thus, Nottale obtained the low-energy asymptotic value of the cosmological constant, which was found to be {\it scale-independent}. Next, he argued that the transition between scale-dependence and scale-independence should be identified with the Thomson scattering length (the classical electron radius), given via $\sigma_{\rm T} \simeq \pi r_{\rm e}^2$, where $\sigma_{\rm T}$ is the scattering cross-section. This is equal to the $e^{+}e^{-}$ annihilation cross-section 
\bea \label{N-5}
\sigma(e^{+}e^{-}) = \pi r_{\rm e}^2\left(\frac{m_{\rm e}c^2}{E}\right) \, , 
\eea
evaluated at $E \simeq m_{\rm e}c^2$. 

In other words, $r_{\rm e}$ represents the effective electron radius, which is an energy-dependent quantity $r(E)$, evaluated at its own mass scale. The cross-section for $e^{+}e^{-}$ pair-production at this energy scale represents the main contribution to the vacuum energy at late times. Hence, by identifying $\rho_{\rm vac} \equiv \rho_{\Lambda}$ and $r \equiv r_{\rm e}$ in Eq. (\ref{N-1}), he obtained the relation
\bea \label{N-6}
\Lambda \simeq \frac{l_{\rm Pl}^4}{r_{\rm e}^6} \, ,
\eea
which is equivalent to Eqs. (\ref{Lambda})-(\ref{r_e}).

%Sec.2.2%%%%%%%%%%%%%%%%%%%%%%%%%%%%%%%%%%%%%%%%%%%%%%%%
\subsection{Dark energy particles and the `Small Number Hypothesis'} \label{sect2.2}

In 2008, Eqs. (\ref{Lambda})-(\ref{r_e}) were also obtained by Boehmer and Harko \cite{Boehmer:2006fd} in 2008, who noticed that the length scale given by the right-hand side of Eq. (\ref{LNH-4}) represents the largest {\it stable} radius of a particle of mass 
\bea \label{m_dS}
m_{\rm dS} \equiv \frac{\hbar}{l_{\rm dS}c} \simeq 10^{-66} {\rm g} \, , 
\eea 
and density $\rho_{\rm min} \simeq \rho_{\Lambda}$. The mass scale $m_{\rm dS}$ was previously proposed by Wesson as the minimum possible mass/energy scale in the present day Universe \cite{Wesson:2003qn}, and the existence of a minimum possible density, $\rho_{\rm min} = \rho_{\Lambda}/2$, for stable, charge-neutral, self-gravitating compact objects in the presence of a positive cosmological constant ($\Lambda >0$), was shown in \cite{Boehmer:2005sm}. This result was obtained directly from the generalised Buchdahl inequalities 
\bea \label{Buchdahl-1}
\frac{2Gm}{c^2} \geq \frac{\Lambda}{6}R^3 \, , \quad \rho = \frac{3m}{4\pi R^3} \geq \rho_{\rm min} \equiv \frac{\Lambda c^2}{16\pi G} \, , 
\eea
where $R$ represents the classical radius of the compact object. Substituting $m_{\rm dS}$ into Eq. (\ref{Buchdahl-1}), we obtain
\bea \label{R}
R \leq (l_{\rm Pl}^2l_{\rm dS})^{1/3} 
%\simeq r_{\rm e} 
\simeq 10^{-13} {\rm cm} \, . 
\eea
Interestingly, it may be shown that this length scale also represents the minimum radius into which the present mass of the Universe, $m_{\rm U} \simeq m'_{\rm ds} \equiv m_{\rm Pl}^2/m_{\rm dS}$, can be compressed without exceeding the Planck density \cite{Burikham:2015nma}.

Noting the numerical coincidence between this length scale and the classical electron radius, Boehmer and Harko proposed a formal equivalence between the two on the basis of a `Small Number Hypothesis' (SNH), which directly yields Eq. (\ref{N-6}). However, since the reciprocal of a large number is a small number, this may be considered logically equivalent to Zel'dovich's reformulation of the LNH for $\Lambda >0$: what differs is the empirical content, according to the substitution $\lambda_{\rm p} \rightarrow r_{\rm e}$. Thus, the work presented in \cite{Boehmer:2006fd,Burikham:2015nma} shows that, not only is the length scale given in Eq. (\ref{R}) comparable to the classical electron radius, it also represents $(i)$ the maximum radius of a minimum-mass, minimum-density particle and $(ii)$ the minimum radius of a maximum-mass, maximum-density particle in the observable Universe. 

However, for charge-neutral, quantum mechanical particles, we may identify the radius $R$ with the Compton wavelength $\lambda_{\rm C} = \hbar/(mc)$ (from here on, we use conventional notation and terminology, referring to the Compton wavelength and reduced Compton wavelength interchangeably), yielding \cite{Burikham:2015nma}
\bea \label{m_Lambda}
m \geq m_{\Lambda} \equiv \frac{1}{\sqrt{2}}\sqrt{m_{\rm Pl}m_{\rm dS}} \simeq 10^{-35} {\rm g} \, .
\eea
Hence, we see that $E_{\rm dS} = m_{\rm dS}c^2$ may be interpreted as the minimum possible quantum of energy -- corresponding to a de Broglie wavelength of the order of the de Sitter horizon -- but {\it cannot} be the minimum rest mass of a stable massive particle. 

The mass scale $m_{\Lambda}$ (\ref{m_Lambda}) has several interesting properties. According to the model presented in \cite{Burikham:2015nma}, it represents the minimum possible mass of a stable, charge-neutral, quantum mechanical and self-gravitating body in the presence of dark energy ($\Lambda >0$). With this in mind, it is notable that it is consistent with current experimental bounds on the mass of the electron neutrino obtained from Planck satellite data \cite{PlanckCollaboration}. It may {\it also} be interpreted as the mass of an effective dark energy particle and its associated Compton wavelength, $l_{\Lambda} = \sqrt{2}\sqrt{l_{\rm Pl}l_{\rm dS}}$, is of the order of $0.1$ mm. Thus, according to this model, the dark energy density is approximately constant over large distances, but may become granular on sub-millimetre scales. In this context, it is notable that that tentative hints of periodic variation in the gravitational field strength on precisely this length scale have recently been observed \cite{Perivolaropoulos:2016ucs}. It is also the {\it unique} mass scale for which the Compton wavelength is equal to the gravitational turn-around radius in the presence of $\Lambda >0$ \cite{Bhattacharya:2016vur}.

As we shall see in Sec. \ref{sect2.4}, $m \simeq m_{\Lambda}$ also arises naturally in the context of the the dark-energy modified uncertainty principle (DE-UP), proposed in \cite{Burikham:2015sro,Lake:2017uzd}, {\it together} with the mass scale $m \simeq m_{\rm e} \simeq \alpha_{\rm e}(m_{\rm Pl}^2m_{\rm dS})^{1/3}$. The former corresponds to an absolute minimum, while the latter is associated with the length scale (\ref{R}) via the relativistic formula for the minimum stable radius of a charged particle, $R \gtrsim Q^2/(mc^2)$, evaluated for $Q = \pm e$.

%Sec.2.3%%%%%%%%%%%%%%%%%%%%%%%%%%%%%%%%%%%%%%%%%%%%%%%%
\subsection{Information theory} \label{sect2.3}

Also in 2008, Eqs. (\ref{Lambda})-(\ref{r_e}) were obtained by Beck using an information theoretic approach to the cosmological constant problem \cite{Beck:2008rd}. He used a system of four axioms, constructed by analogy with the Kinchin axioms of information theory \cite{Khinchin}, which describe the most `desirable' properties of an information measure, to fix the form of the cosmological constant in terms of the other fundamental constants of nature. 

It may be shown that the Kinchin axioms uniquely fix the form of the Shannon entropy, which forms the mathematical, though not the microphysical, basis of statistical mechanics and thermodynamics. Thus, by constructing an analogous approach to the cosmological constant problem, Beck attempted to fix the form of $\Lambda$ in terms of the other constants of nature on an axiomatic basis, without reference to an underlying microphysical theory.    

By formally replacing the dependence of the information measure $I$ on the probabilities of events $p_i$ by the dependence of $\Lambda$ on the remaining physical constants, i.e., the fundamental coupling constants $\alpha_i$, masses $m_i$ and mixing angles $s_i$, he argued that the requirements of Fundamentality (L1), Boundedness (L2), Simplicity (L3) and Scale-invariance (L4) uniquely fix the form of the cosmological constant according to Eqs. (\ref{Lambda})-(\ref{r_e}). These axioms are constructed by analogy with the four Kinchin axioms (K1-K4) of the same names.

Specifically, K1, `fundamentality', simply states that the information measure $I$ should depend on the fundamental quantities, the probabilities of events $p_i$,  
\bea \label{K1}
I = I(p_i) \, , \quad ({\rm K1})  
\eea
and not on any other factors. The analogous axiom L1 for the cosmological constant is
\bea \label{L1}
\Lambda = \Lambda(\left\{\alpha_i\right\},\left\{m_i\right\},\left\{s_i\right\}) \, , \quad ({\rm L1}) \, .   
\eea
K2, `boundedness' states that there exists a lower bound for the value of $I$, corresponding to the uniform distribution, $p_i = 1/N$, where $N$ is the total number of distinct events, such that
\bea \label{K2}
I(1/N,1/N . . . 1/N) \leq I(p_1,p_2 . . . p_{N}) \, , \quad ({\rm K2})  
\eea
The analogous axiom L2 states
\bea \label{L2}
0 < \Lambda \, , \quad ({\rm L2})  
\eea
where $\Lambda = 0$ is explicitly excluded. K3, `simplicity', states that the information measure should not change if the set of events is enlarged by another set with probability zero, i.e.
\bea \label{K3}
I(p_1,p_2 . . . p_{N}) = I(p_1,p_2 . . . p_{N};0) \, , \quad ({\rm K3}) \, .   
\eea
The analogous axiom L3 is
\bea \label{L3}
\Lambda(\left\{\alpha_i\right\},\left\{m_i\right\},\left\{s_i\right\}) = \Lambda(\left\{\alpha_i\right\},\left\{m_i\right\},\left\{s_i\right\};\left\{c_i\right\})
%\, , \,
({\rm L3}) 
%\, ,
\eea
where the $c_i$ are {\it not} fundamental constants of nature. The final axiom, K4 `invariance' is the most restrictive. It may be expressed as 
\bea \label{}
I\left(\left\{p_{ij}^{I,II}\right\}\right) = I(\left\{p_{i}^{I}\right\}) + \sum_{i} p_{i}^{I}I\left\{(p^{II}(j|i)\right\}) \, ,
\eea
where the superscripts $I$ and $II$ denote probabilities of events in different  (not necessarily independent) subsystems and $p^{II}(j|i)$ is the conditional probability of event $j$ in subsystem $II$, given an event $i$ in subsystem $I$.
$I(\left\{p^{II}(j|i)\right\})$ is the conditional information of subsystem $II$, in the joint system $I,II$, described by the probabilities $p_{ij}^{I,II} = p_i^{I}p^{II}(j|i)$. The meaning of K4 is that the information measure should be independent of the way in which the information is collected. We may either (a) collect information in subsystem $I$, then in subsystem $II$, or (b) collect information in subsystem $II$, {\it assuming} a given event in subsystem $I$, before summing over all possible events in subsystem $I$, weighted by their respective probabilities $p_i$. Hence, there is a {\it scale transformation} in the space of possible information measures and probabilities, such that
\bea \label{K4}
I(\left\{\tilde{p}_i\right\}) = \tilde{I}(\left\{p_i\right\}) \, , \quad ({\rm K4}) \, ,
\eea
where a tilde denotes transformed quantities. The analogous axiom L4 may therefore be expressed as 
\bea \label{L4}
\Lambda(\left\{\tilde{\alpha}_i\right\},\left\{\tilde{m}_i\right\},\left\{\tilde{s}_i\right\}) = \tilde{\Lambda}(\left\{\alpha_i\right\},\left\{m_i\right\},\left\{s_i\right\}) \, , \quad ({\rm K4}) \, ,
\eea
where a tilde also denotes an appropriate scale transformation: in this case, a literal rescaling of the fundamental constants of nature by an arbitrary numerical factor. 

After formulating the axioms L1-L4, the argument presented in \cite{Beck:2008rd} relies on three specific assumptions, $(i)$ gravitational scale invariance, $(ii)$ a dimensional argument and $(iii)$ electromagnetic scale invariance. The first and last of these are implicit in axiom L4, but Beck singled out the electromagnetic and gravitational interactions as being of greatest relevance to the large-scale dynamics of the early (before recombination) and late-time (after recombination) Universe, respectively. We now consider each of these assumptions in detail.

$(i)$ Gravitational scale-invariance: In the Newtonian approximation, the the gravitational energy density of a distribution of point-like masses occupying a volume $V$ is
\bea \label{gravE}
\rho_{G} = -\frac{G}{V}\sum_{i,j}\frac{m_im_j}{r_{ij}} \, ,
\eea
where $r_{ij}$ denotes the distance between the $i^{\rm th}$ and $j^{\rm th}$ masses. Thus, if the gravitational constant $G$ is rescaled such that
\bea
G \rightarrow \Gamma G \, ,
\eea
where $\Gamma$ is an arbitrary numerical constant, but the masses $m_i$, $m_j$ and distances $r_{ij}$ are kept the same, the energy density scales as
\bea
\rho_{G} \rightarrow \Gamma \rho_{G}  \, .
\eea
Scale invariance of the ratio $\rho_{\rm vac}/\rho_{G}$ then requires 
\bea
\rho_{\rm vac} \rightarrow \Gamma \rho_{\rm vac}  \, .
\eea
Hence, $\rho_{\rm vac} \propto G$ and we may set
\bea
\rho_{\rm vac} \sim GX \, , 
\eea
where $X$ is an, as yet unknown, quantity.

$(ii)$ The dimensional argument: On purely dimensional grounds, the unknown factor $X$ must take the form $X \sim (c^4/\hbar^4)m^6$, where $m$ is an arbitrary mass scale, composed (in some way) from fundamental mass scales, dimensionless coupling constants and mixing angles, according to axiom L1. (Note that this dimensional argument is unaffected even if the fundamental constants that are {\it not} themselves fundamental couplings, masses or mixing angles, i.e. $\hbar$ and $c$, are also rescaled by $\Gamma$.) Thus, without loss of generality, we may set
\bea
\rho_{\rm vac} = A\frac{c^4}{\hbar^4}m_{\rm e}^6 \, , 
\eea
where $A$ is a numerical constant, which may depend on the remaining constants of nature $\left\{\left\{\alpha_i\right\},\left\{m_i\right\},\left\{s_i\right\}\right\}$ in any way. Likewise, $A$ may be expressed arbitrarily in terms of $\alpha_{\rm e}$, such that
\bea \label{8pi}
\rho_{\rm vac} = \frac{1}{8\pi}\alpha_{\rm e}^{\eta}\frac{c^4}{\hbar^4}m_{\rm e}^6 \, , 
\eea
where $\eta$ is an arbitrary function of $\left\{\left\{\alpha_i\right\},\left\{m_i\right\},\left\{s_i\right\}\right\}$. 

$(iii)$ Electromagnetic scale-invariance: Beck then argued that the most important process determining the large-scale evolution of the Universe before recombination was the Thomson scattering of electrons, associated with the scattering length $r_{\rm e} \sim \sqrt{\sigma_{\rm T}}$. Clearly, this line of reasoning is {\it similar} to Nottale's and Beck could also have argued {\it {\` a} la} Nottale that $r_{\rm e}$, interpreted as the effective cross-sectional radius for $e^{+}e^{-}$ pair-production, is just as relevant to the large-scale dynamics of the late-time Universe. Finally, therefore, Beck claimed that, since simultaneous scale transformations of the form
\bea
\alpha_{\rm e} \rightarrow \Gamma \alpha_{\rm e}  \, , \quad m_{\rm e} \rightarrow \Gamma m_{\rm e}  \, ,
\eea 
leave the Thomson scattering cross-section $\sigma_{\rm T}$ invariant, scale-invariance of the ratio $\Lambda^{-1}/\sigma_{\rm T}$ requires that $\Lambda$ be dependent on the ratio $m_{\rm e}/\alpha_{\rm e}$ {\it only}. This fixes the remaining free constant, $\eta = -6$, and the factor of $(8\pi)^{-1}$ in Eq. (\ref{8pi}) is chosen so as to match the numerical factor in Eq. (\ref{rho_Lambda}), giving the `simplest' result:
\bea
\rho_{\Lambda} = \frac{\Lambda c^2}{8\pi G} \equiv \rho_{\rm vac} = \frac{1}{8\pi}\frac{c^4}{\hbar^4}\left(\frac{m_{\rm e}}{\alpha_{\rm e}}\right)^{6} \, . 
\eea
It is straightforward to check that this is relation is equivalent to Eqs. (\ref{Lambda})-(\ref{r_e}) and (\ref{N-6}).

However, it must be noted that an obvious problem with Beck's approach is that, unlike the probabilities of events, which are {\it all} dimensionless, the fundamental quantities on which he bases the expression for the cosmological constant $\Lambda$ are inequivalent in this respect. Though {\it non}-gravitational coupling constants may be expressed in dimensionless form using $\hbar$ and $c$, Newton's constant $G$ cannot. Similarly, the fundamental masses cannot be expressed in dimensionless form without the use of the Planck mass, $m_{\rm Pl} \propto 1/{\sqrt{G}}$. A related problem concerns the fact that, if the ultimate origin of the point-like masses considered in Eq. (\ref{gravE}) are {\it fundamental} masses, these should also be rescaled at the same time as $G$, which changes the scaling of $\rho_{\rm vac}$ with $\Gamma$, dramatically. 

Overall, one can say that it does not make physical sense to rescale {\it dissimilar} quantities (i.e., those with inequivalent dimensions) with the same scale factor: only similar quantities should be rescaled in this way. A self-consistent set of rescalings, which leave dimensionless quantities constructed from Eq. (\ref{N-6}) explicitly invariant (in line with axiom L4), would then be
\bea \label{rescale}
\Lambda^{-1} \rightarrow \Gamma \Lambda^{-1} \, , \quad \sigma_{\rm T} \rightarrow \Gamma \sigma_{\rm T} \, , \quad l_{\rm Pl}^2 \rightarrow \Gamma l_{\rm Pl}^2 \, .
\eea

Alternatively, instead of requiring the absolute scale independence of $\Lambda$, or even the relative scale independence suggested by Eq. (\ref{rescale}), we may instead require {\it holographic} relationships between the bulk and the boundary of the Universe to be preserved by our expression for $\Lambda$, written in terms of the remaining constants of nature. In the following section, we summarize the results of a new approach to deriving the expressions (\ref{Lambda})-(\ref{r_e}) and (\ref{N-6}), which automatically implements the holographic principle \cite{tHooft:1999rgb,Bousso:2002ju}. 

%Sec.2.4%%%%%%%%%%%%%%%%%%%%%%%%%%%%%%%%%%%%%%%%%%%%%%%%
\subsection{Minimum length uncertainty relations in a dark energy Universe} \label{sect2.4}

A further attempt to explain the connection between cosmological and atomic scales in the LNH, in terms of the stability of fundamental particles in the presence of dark energy, was presented in a series of recent papers \cite{Burikham:2015sro,Lake:2017uzd}. (See also \cite{Burikham:2016cwz,Burikham:2017bkn} for extensions to modified gravity theories.) This approach considered the status of minimum length uncertainty relations (MLURs), motivated by quantum gravity phenomenology (see \cite{Hossenfelder:2012jw,Garay:1994en} for contemporary reviews), in the presence of a vacuum energy density $\rho_{\Lambda}$ given by Eq. (\ref{rho_Lambda}). 

In \cite{Burikham:2015sro}, a new form of MLUR, dubbed the `dark energy uncertainty principle' or DE-UP for short, which explicitly includes the de Sitter scale $l_{\rm dS} \sim 1/\sqrt{\Lambda}$, as well as the Planck scale, was proposed. The general form of the DE-UP proposed in \cite{Burikham:2015sro} is
\begin{eqnarray} \label{gen}
\Delta x_{\rm total}(\Delta v,r,m) &=& \Delta x_{\rm canon.}(\Delta v,r,m) + \Delta x_{\rm grav}(r,m)  
\nonumber\\ 
&\geq& \Delta x(\Delta v) + \Delta x_{\rm recoil}(\Delta v,r,m) 
\nonumber\\ 
&+& \Delta x_{\rm grav}(r,m) 
\nonumber\\ 
&\geq&  (\Delta x_{\rm canon.})_{\rm min}(r,m) 
\nonumber\\ 
&+& \Delta x_{\rm grav}(r,m) \, .
\end{eqnarray}
Here, $(\Delta x_{\rm canon.})_{\rm min}$ denotes the minimum possible canonical quantum uncertainty of a wave packet that has been freely evolving for a time $t = r/c$. By explicitly solving the free-particle Schr{\"o}dinger equation in the Heisenberg picture, it is straightforward to show that, for a particle of mass $m$, the minimum canonical uncertainty is 
\bea \label{canon_min}
(\Delta x_{\rm canon.})_{\rm min} \simeq \sqrt{\lambda_{\rm C}r} \, , 
\eea
where $r=ct$ and $\lambda_{\rm C}$ is the Compton wavelength \cite{Calmet:2004mp,Calmet:2005mh}. This expression can also be obtained by considering a gedanken experiment, originally due to Salecker and Wigner, in which a massive particle is used to `measure' a distance $r \simeq ct$ by means of the emission and reabsorption of a photon  \cite{Salecker:1957be}. By minimising the sum of the first two terms on the top line of Eq. (\ref{gen}), where $\Delta x$ represents the canonical Heisenberg uncertainty and $\Delta x_{\rm recoil} \simeq \Delta vt \equiv \Delta p r/(mc)$ is the additional uncertainty due to recoil, with respect to either $\Delta v$ or $m$, we obtain Eq. (\ref{canon_min}) directly. 

The term $\Delta x_{\rm grav}$ represents an additional contribution to the total uncertainty, due to the superposition of gravitational field states which are (in turn) induced by the superposition of position states associated with $m$. From a relativistic perspective, this may be considered {\it equivalent} to the uncertainty associated with the superposition of space-time geometries in the quantum gravity regime. In \cite{Burikham:2015sro}, it was conjectured that this superposition is influenced by two factors: $(i)$ the mass of the particle, and $(ii)$ the presence of the dark energy density, or, equivalently, of a finite horizon for the wave function centre of mass, $l_{\rm dS} \sim 1/\sqrt{\Lambda}$. Taking both these factors into account, the conjectured form of $\Delta x_{\rm grav}$ proposed in \cite{Burikham:2015sro} was
\bea \label{grav}
\Delta x_{\rm grav} \simeq \frac{1}{\sqrt{\Lambda}}\frac{Gm}{c^2r} \simeq \frac{l_{\rm Pl}^2l_{\rm dS}}{\lambda_{\rm C}r} \, . 
\eea

This was combined with the limit on the mass/radius ratio for stable, charged, compact objects originally obtained by Bekenstein \cite{Bekenstein:1971ej}, and later generalised by Boehmer and Harko for $\Lambda >0$ \cite{Boehmer:2007gq}, i.e.
\bea \label{Bek}
R \gtrsim \frac{Q^2}{mc^2} + {\rm h.o.t.}(G,\Lambda \ . \ . \ ) \, . 
\eea 
To leading order, Eq. (\ref{Bek}) simply recovers the well-known formula for the classical radius of a charged particle obtained from special relativity, but this remains rigorously valid in the weak-field limit of general relativity, even in the presence of dark energy \cite{Bekenstein:1971ej,Boehmer:2007gq}. By identifying the {\it total} minimum quantum mechanical uncertainty $(\Delta x_{\rm total})_{\rm min}$, including both canonical quantum and gravitational/dark energy-induced terms, with the minimum radius $R$ obtained from the Bekenstein bound, the following inequality is obtained 
\bea \label{bound}
\frac{Q^2}{mc^2} \lesssim (l_{\rm Pl}^2l_{\rm dS})^{1/3} \, . 
\eea 
Thus, according to the model presented in \cite{Burikham:2015sro}, Eq. (\ref{bound}) gives the maximum possible charge-squared to mass ratio for a stable, charged, self-gravitating and quantum mechanical object. Assuming saturation of this bound is equivalent to assuming the existence of a particle in nature that simultaneously saturates both the classical (general-relativistic) {\it and} quantum mechanical (MLUR) stability constraints. Remarkably, comparison with Eq. (\ref{r_e}) reveals that saturation is obtained for the electron charge-squared to mass ratio. Equivalently, we see that evaluating (\ref{bound}) for $Q = \pm e$, yields
\begin{eqnarray}  \label{me}
&& m \gtrsim \alpha_{e}(m_{\rm Pl}^2m_{\rm dS})^{1/3} = 7.332 \times 10^{-28} \, {\rm g}
\nonumber\\
&\simeq& m_{e} = 9.109 \times 10^{-28} \, {\rm g} \, .
\end{eqnarray}
Hence, if the electron were more highly charged (with the same mass $m_{\rm e}$) or any less massive (with the same charge $e$), a combination of electrostatic and dark energy repulsion would destabilize its Compton wavelength \cite{Burikham:2015sro}.

In \cite{Lake:2017uzd}, a full {\it physical} derivation of the dark energy-modified uncertainty principle (DE-UP), proposed in \cite{Burikham:2015sro} is given, in which it is shown that the de Sitter length-dependent term arises as a direct consequence of the existence of a finite horizon $r_{\rm H}(\tau_0) \simeq l_{\rm dS}$, where $\tau_0$ is the present age of the Universe. In addition, it is shown that the DE-UP defined by Eqs. (\ref{gen}) and (\ref{canon_min})-(\ref{grav}) naturally incorporates the mass bound for neutral particles, Eq. (\ref{m_Lambda}), in addition to that for charged particles, Eq. (\ref{me}). In particular, it is straightforward to show that requiring every potentially observable length-scale, i.e. $(\Delta x_{\rm canon.})_{\rm min}$, $\Delta x_{\rm grav}$ and the `probe' distance $r$, to be super-Planckian, automatically implies the existence of a minimum mass in nature, $m_{\Lambda} \simeq \sqrt{m_{\rm Pl}m_{\rm dS}}$. Alternatively, beginning with the bound (\ref{m_Lambda}), obtained by combining both general-relativistic and quantum mechanical effects, we obtain $l_{\rm Pl}$ as a limiting resolution for physical measurements of length within the DE-UP framework.

Finally, we note that, using Salam's theory of strong gravity \cite{Isham:1971gm,Isham:1970aw,Isham:1973jh,Isham:1974wx} as an {\it effective} theory for modeling quark confinement, analogous arguments where applied to charged, strongly interacting particles, in which the energy density associated with the `strong de Sitter radius' was identified with the `bag constant' of the MIT bag model for confined nuclear matter, $B \simeq 10^{14} \ {\rm gcm^{-1}}$ \cite{Burikham:2017bkn}. These arguments successfully predicted the correct order of magnitude value for the charge-squared to mass ratio of the up quark, the lightest known strongly interacting and quantum mechanical particle, with charge $Q = \pm 2e/3$ \cite{Burikham:2017bkn}.

%Sec.3%%%%%%%%%%%%%%%%%%%%%%%%%%%%%%%%%%%%%%%%%%%%%%%%
%%%%%%%%%%%%%%%%%%%%%%%%%%%%%%%%%%%%%%%%%%%%%%%%%%%%
\section{Implications of a connection between ``dark" and ``light" physics} 
%-- holography, Universal expansion, and directions for future research} 
\label{sect3}

%Sec.3.1%%%%%%%%%%%%%%%%%%%%%%%%%%%%%%%%%%%%%%%%%%%%%%%%
\subsection{Holography} \label{sect3.1}

It is straightforward to see that, for any particle that minimizes the total uncertainty given by the DE-UP, Eqs. (\ref{gen}) and (\ref{canon_min})-(\ref{grav}), a holographic relation holds between the bulk and the boundary of the Universe, namely
\begin{eqnarray}  \label{holog}
\left(\frac{(\Delta x_{\rm total})_{\rm min}}{l_{\rm dS}}\right)^3 = \frac{l_{\rm Pl}^2}{l_{\rm dS}^2} = N = 1.030 \times 10^{122} \, .
\end{eqnarray}
Hence, the number of Planck sized `bits' on the de Sitter boundary is equal to the number of minimum-volume `cells', $V_{\rm cell} \sim (\Delta x_{\rm total})_{\rm min}^3$, in the bulk \cite{Burikham:2015sro}. As mentioned previously, in Sec. \ref{sect2.2}, $R \simeq (\Delta x_{\rm total})_{\rm min}$ may also be interpreted as the {\it classical} radius of a `particle' with both minimum energy, $E_{\rm dS} = m_{\rm dS}c^2$, and minimum density, $\rho_{\rm min} \simeq \rho_{\Lambda}/2$, i.e.
\begin{eqnarray}  \label{}
&&\rho \simeq \frac{m_{\rm dS}}{R^3} \gtrsim \rho_{\rm min} \simeq \frac{\Lambda c^2}{G} %\simeq \frac{m_{\rm Pl}}{l_{\rm dS}^2l_{\rm Pl}} 
\nonumber\\
&\iff& R \lesssim (\Delta x_{\rm total})_{\rm min} \simeq (l_{\rm Pl}^2l_{\rm dS})^{1/3} \, . 
\end{eqnarray}
Although a massive particle with rest energy $E_{\rm dS}$ would be unstable due to the repulsive effects of dark energy, this {\it may} correspond to the energy of a photon with maximum wavelength, $\lambda \simeq l_{\rm dS}$. Thus, $(\Delta x_{\rm total})_{\rm min}$ may also be interpreted as the classical radius of a localized, minimum-energy photon. A space-filling `sea' of such photons would have the same energy density as the dark energy field \cite{Burikham:2015nma}. 

In addition, we may consider a maximum-mass, maximum-density state, for which $\rho \simeq \rho_{\rm Pl}$ and the total energy is $E_{\rm dS}' \simeq m_{\rm dS}'c^2$ ($m'_{\rm dS} \equiv m_{\rm Pl}^2/m_{\rm dS}$). The classical radius thus obtained corresponds to the smallest possible volume within which the total mass of the present day horizon may be confined, without exceeding the Planck density. We then have \cite{Burikham:2015sro}
\begin{eqnarray}  \label{}
&&\rho \simeq \frac{m_{\rm dS}'}{R^3} \lesssim \rho_{\rm Pl} \simeq \frac{c^5}{\hbar G^2} 
\nonumber\\
&\iff& R \gtrsim (\Delta x_{\rm total})_{\rm min} \simeq (l_{\rm Pl}^2l_{\rm dS})^{1/3} \, . 
\end{eqnarray}
%As stated in Sec. \ref{sect2.2}, 
The length scale $(\Delta x_{\rm total})_{\rm min} \simeq (l_{\rm Pl}^2l_{\rm dS})^{1/3}$ therefore corresponds to at least three interesting physical scenarios. It may be interpreted as $(i)$ the classical radius of a `particle' with both minimum energy and minimum energy density, $(ii)$ the classical radius of a `particle' with both maximum energy and maximum energy density, and $(iii)$ the classical radius/minimum total uncertainty of the electron, which saturates the charged particle stability bound (\ref{bound}). In the context of the DE-UP model \cite{Burikham:2015sro,Lake:2017uzd}, {\it all three} interpretations satisfy the general holographic relation, Eq. (\ref{holog}).

%Sec.3.2%%%%%%%%%%%%%%%%%%%%%%%%%%%%%%%%%%%%%%%%%%%%%%%%
\subsection{Universal expansion} \label{sect3.2}

As pointed out in \cite{Burikham:2015nma}, the minimum mass $m_{\Lambda} \simeq \sqrt{m_{\rm Pl}m_{\rm dS}}$ (\ref{m_Lambda}) may also be interpreted as the mass of an {\it effective} dark energy particle. Since, in this model, even random quantum fluctuations reduce the inter-particle distance between nearest neighbours to less than $\lambda_{\rm C}(m_{\Lambda}) = l_{\Lambda} \simeq \sqrt{l_{\rm Pl}l_{\rm dS}}$, it follows that the pair-production of ``dark'' minimum-mass particles is capable of driving the present day accelerated expansion of the Universe. In short, if space is `full'  of dark energy particles, with mean inter-particle distance $l_{\Lambda} \sim 0.1$ mm, the pair-production necessitated by quantum mechanics {\it requires} a concomitant expansion of space \cite{Burikham:2015nma,Burikham:2017bkn}.

Let us assume that the probability of a single (holographic) spatial cell `pair-producing' within a time interval $\Delta \tau = t_{\rm Pl} = l_{\rm Pl}/c$, due to the pair-production of dark energy particles, is given by
\begin{eqnarray}  \label{}
&&P(\Delta V = +V_{\rm cell}|V_0=V_{\rm cell},\Delta \tau =t_{\rm Pl}) = N^{-1/2}
\nonumber\\ 
&& = \frac{V_{\rm Pl}}{V_{\rm cell}} = \frac{l_{\rm Pl}}{l_{\rm dS}} \simeq \left(\frac{\hbar G \Lambda}{3c^3}\right)^{1/2} \simeq 9.851 \times 10^{-62} \, ,
\end{eqnarray}
where $V_0$ denotes the initial volume at the initial time. This leads naturally to a de Sitter-type expansion, modelled by the differential equation 
\begin{eqnarray} \label{dS_exp-1}
\frac{d a^3}{d\tau} \simeq \frac{N^{-1/2} a^3}{t_{\rm Pl}} = \frac{l_{\rm Pl}}{l_{\rm dS}}\frac{a^3}{t_{\rm Pl}} \, ,
\end{eqnarray}
or, equivalently, 
\begin{eqnarray}\label{dS_exp-2}
\frac{d a}{d\tau} \simeq c\sqrt{\frac{\Lambda}{3}}a \, ,  \quad a(\tau) \simeq a_0e^{-c\sqrt{\Lambda/3}\tau} \, .
\end{eqnarray}
Since the production of a single dark energy energy particle requires the production of $n_{\rm cell} = V_{\Lambda}/V_{\rm cell} \simeq l_{\Lambda}^3/(l_{\rm Pl}^2l_{\rm ds}) = N^{1/4}$ cells of space, this implies that the probability of a dark energy particle pair-producing within a single Planck time is given by
\begin{eqnarray}  \label{}
P(\Delta V &=& +V_{\rm \Lambda}|V_0=V_{\rm \Lambda},\Delta t =t_{\rm Pl}) \simeq N^{-3/4} 
\nonumber\\
&=& \left(\frac{l_{\rm ds}}{l_{\rm Pl}}\right)^{-3/2} \simeq 10^{-91} \, .
\end{eqnarray}
However, since there are $n_{\rm DE} \simeq l_{\rm dS}^3/l_{\Lambda}^3 = N^{3/4}$ dark energy particles within the de Sitter horizon, this means that one dark energy particle is produced, {\it somewhere} in the observable Universe, during every Planck time interval. This rate of pair-production is capable of giving rise to the accelerated Universal expansion observed at the present epoch. 

In this model, the observed vacuum energy is really the energy associated with the dark energy field, for which $\lambda_{\rm C}(m_{\Lambda}) = l_{\Lambda}$ provides a natural a UV cut-off for the field modes, yielding
\begin{eqnarray} \label{rho_vac}
\rho_{\rm vac} \simeq \frac{\hbar}{c} \int_{1/l_{\rm dS}}^{1/l_{\Lambda}} \sqrt{k^2 + \left(\frac{2\pi}{l_{\Lambda}}\right)^2} {\rm d}^3k 
\nonumber\\
\simeq \frac{m_{\rm Pl}l_{\rm Pl}}{l_{\Lambda}^{4}} \simeq \frac{\Lambda c^2}{G} \simeq 10^{30} \ {\rm gcm^{-3}} \, .
\end{eqnarray}
The field itself remains `trapped' in a Hagedorn-type phase, in which any increase in kinetic energy, {\it even that caused by random collisions between neighbouring dark energy particles due to quantum uncertainty}, results in pair-production, rather than an increase in temperature. 

The temperature associated with the field therefore remains constant, on large scales, and is comparable to the present day temperature of the CMB,
\begin{eqnarray} \label{T_Lambda}
T_{\Lambda} \equiv \frac{m_{\Lambda}c^2}{8 \pi k_{\rm B}} \simeq 2.27 \ {\rm K} \simeq T_{\rm CMB} = 2.73 \ {\rm K} \, . 
\end{eqnarray}
Here the factor of $8\pi$ is included by analogy with the expression for the Hawking temperature, 
\begin{eqnarray} \label{T_H}
T_{\rm H} \equiv \frac{c^2}{8 \pi k_{\rm B}}\frac{m_{\rm Pl}^2}{m} \, ,
\end{eqnarray}
so that $T_{\Lambda} \equiv T(m_{\Lambda}) = T_{\rm H}(m_{\Lambda}')$, where $m_{\Lambda}' \equiv m_{\rm Pl}^2/m_{\Lambda}$ is the dual mass. Though this may seem like another `miraculous' coincidence, {\it {\`a} la} Dirac, in the dark energy model implied by the DE-UP it is simply a re-statement of the standard `coincidence problem' of cosmology, i.e. the Universe begins a phase of accelerated expansion when $r_{\rm U} \simeq l_{\rm dS}$, at which point $\Omega_{\rm M} \simeq \Omega_{\Lambda}$ and, hence, $T_{\rm CMB} \simeq T_{\Lambda}$. The question remains, why do we live at precisely this epoch? However, no {\it new} coincidences are required, in order to `explain' Eq. (\ref{T_Lambda}).

Again, we may apply analogous arguments to strongly interacting particles by using Salam's theory of strong gravity as an {\it effective} theory for confined (and deconfined) nuclear matter. These predict a genuine Hagedorn temperature for the quark-gluon plasma of order $T_{\rm Hag} \simeq N^{1/10}T_{\Lambda} \simeq 10^{12}$ K \cite{Burikham:2017bkn}.

%Sec.3.3%%%%%%%%%%%%%%%%%%%%%%%%%%%%%%%%%%%%%%%%%%%%%%%%
\subsection{Additional cosmological implications} \label{sect3.3}

The holographic relation Eq. (\ref{holog}) remains valid for all epochs, prior to the present day, under the substitution $l_{\rm dS} \rightarrow r_{\rm H}(\tau)$, where $r_{\rm H}(\tau)$ is the physical horizon radius at cosmic time $\tau$. However, under these circumstances, we note that the DE-UP model naturally implies both a time-dependent minimum mass for neutral particles {\it and} a time-dependent maximum charge-squared to mass ratio for charged particles, i.e.
\begin{eqnarray}  \label{neutral}
m_{\rm \nu}(\tau) \gtrsim \sqrt{m_{\rm Pl}m_{\rm H}(\tau)} \, ,
\end{eqnarray}
where $m_{\rm H}(\tau) \equiv \hbar/(r_{\rm H}(\tau)c)$, and
\begin{eqnarray}  \label{charged}
\frac{e^2}{m_e}(\tau) 
%\lesssim \left(\hbar^2G^2c^6 r_{\rm H}^2(\tau)\right)^{1/6} 
\lesssim c^2(l_{\rm Pl}^2r_{\rm H}(\tau))^{1/3}\, ,
\end{eqnarray}
respectively \cite{Lake:2017uzd}. Equation (\ref{charged}) corresponds to a three-dimensional, time-dependent, holographic cell radius
\begin{eqnarray}  \label{cell}
(\Delta x_{\rm total})_{\rm min}(\tau) \simeq (l_{\rm Pl}^2r_{\rm H}(\tau))^{1/3} \, .
\end{eqnarray}
This is similar to the MLUR for an expanding Universe suggested by Ng \cite{Ng:2016wzj}, but with the cosmological horizon $r_{\rm H}(\tau)$ in place of the Hubble horizon $\mathcal{H}(\tau)/c$. 

Though highly speculative, these relations imply an interesting form of `unification' in the early Universe, with all masses tending to the Planck mass and all charges tending to the Planck charge, $q_{\rm Pl} = \sqrt{\hbar c}$, as $r_{\rm H}(\tau) \rightarrow l_{\rm Pl}$. Alternatively, if the Planck-density threshold limits the radius of the `initial' big bang horizon such that $r_{\rm H}(\tau) \gtrsim (l_{\rm Pl}^2l_{\rm dS})^{1/3}$, as suggested by the results obtained in \cite{Burikham:2015sro}, the minimum holographic cell radius will be of order
\bea
R_{\rm min} \simeq  (l_{\rm Pl}^8l_{\rm dS})^{1/9} \simeq 10^{-26} \ {\rm cm} \, .
%10^{-28} \ {\rm m} \, .
\eea
Finally, we note that the length and time scales associated with the mass $m_{\rm \nu}(\tau)$ (\ref{neutral}), i.e.
\bea
r_{\rm \nu}(\tau) = ct_{\rm \nu}(\tau) \equiv \frac{\hbar}{m_{\rm \nu}(\tau)c} \, , 
\eea
also satisfy the `four-dimensional' holographic relation
\bea \label{holog4D}
\left(\frac{r_{\rm H}(\tau)}{r_{\rm \nu}(\tau)}\right)^4 = \frac{r_{\rm H}^2(\tau)}{l_{\rm Pl}^2} = N(\tau) \lesssim 1.030 \times 10^{122} \, .
\eea

Hence, the DE-UP model strongly suggests time-variation of either, or both, $e$ and $m_{\rm e}$, assuming that $\left\{G,c,\hbar,\Lambda \right\}$ are genuine universal constants. Similar arguments apply to the mass of the lightest neutral particle, previously identified with the mass of the electron neutrino, $m_{\nu}$. 

For models involving temporal and/or spatial variation of fundamental constants, the situation is even more complicated, and it may be extremely difficult, in practice, to distinguish variation in $e$ and/or $m_{\rm e}$, and $m_{\nu}$, from other effects. Particular classes of models, in which the variation of physical constants should, automatically, imply a modification of the DE-UP formulae, Eqs. (\ref{gen}) and (\ref{canon_min})-(\ref{grav}), include those with a running gravitational coupling \cite{Gomez-Valent:2016nzf,Fritzsch:2016ewd,Sola:2016jky,Sola:2016vis}, variable speed of light \cite{Albrecht:1998ir,Barrow:1998eh,Albrecht:1998ij,Moffat:2002nm,Qi:2014zja,Landau:2000mx,Magueijo:2003gj,Dabrowski:2016lot}, or dynamical dark energy field \cite{Tsujikawa:2010zz,AmendolaTsujikawa(2010),Li:2013hia,Li:2012dt}. (See also \cite{Barrow:1998df,Barrow:1999st,Moffat:2001sf,Kozlov:2008iy,Barrow:2014vva} for current bounds on varying $\alpha_e$ theories, including their effects on cosmic string phenomenology \cite{Lake:2010wt,Lake:2010qsa} and \cite{Brandenberger:1999bi,Gopakumar:2000kp,Biesiada:2003tp,Barrow:2005sv,Scoccola:2009zh} for more general models involving variations of multiple physical constants.) 

In fact, several models incorporating non-minimal couplings between dark energy and the electromagnetic sector have already been proposed in the literature, as solutions to problems in contemporary cosmology \cite{Yan:2010nx,Narimani:2011rb,Hollenstein:2012mb,Pandolfi:2014jka,Martins:2015jta}. Although a thorough analysis of the cosmological implications of the DE-UP has not yet been attempted, the cosmological implications of $\Lambda \propto \alpha_{\rm e}^{-6}$ cosmology were investigated in \cite{Wei:2016moy,Wei:2017mzf}, in the context of a time-varying fine structure constant. Nonetheless, its unusual predictions suggest that future observations and/or analysis of currently available data may be capable of falsifying the model  \cite{Lake:2017uzd}.

%Sec.4%%%%%%%%%%%%%%%%%%%%%%%%%%%%%%%%%%%%%%%%%%%%%%%%
%%%%%%%%%%%%%%%%%%%%%%%%%%%%%%%%%%%%%%%%%%%%%%%%%%%%
\section{Discussion} \label{sect4}

We have shown that the relations (\ref{Lambda})-(\ref{r_e}) and (\ref{N-6}), which are equivalent to Zel'dovich's reformulation of Dirac's Large Number Hypothesis for a Universe with $\Lambda >0$, under the identification $m \simeq m_{\rm e}/\alpha_{\rm e}$, are well motivated from a number of theoretical perspectives. Each of these goes beyond numerology and aims to base the seemingly incredible coincidences noted by Dirac, Zel'dovich and others on firm {\it physical} arguments. Specifically, we have outlined four sets of independent, yet not necessarily incompatible arguments, given in the literature, which give rise to the relation $\Lambda \simeq l_{\rm Pl}^4/r_{\rm e}^{6}$ (\ref{N-6}), where $r_{\rm e} = e^2/(m_{\rm e}c^2)$ is the classical electron radius. 

The first, proposed by Nottale \cite{Nottale} (1993), is based on the assumption that the vacuum energy density is dominated by the gravitational energy associated with $e^{+}e^{-}$ pair production. By perturbatively expanding the renormalisation group equation for $\rho_{\rm vac}$, and identifying the transition from running to scale-independence with the energy scale $E \simeq m_{\rm e}c^2$, he obtained Eq. (\ref{N-6}) directly.

The second, proposed by Boehmer and Harko \cite{Boehmer:2006fd} (2008), identifies the maximum stable radius of a minimum-mass, minimum-density `particle', in the presence of a positive cosmological constant $\Lambda > 0$, with the classical electron radius via a `Small Number Hypothesis'. Since the reciprocal of a large number is a small number, this is logically equivalent to Zel'dovich's reformulation of the LNH with $m \simeq m_{\rm e}/\alpha_{\rm e}$. However, the important {\it physical} content of this work is the realisation that the cosmological constant automatically implies the existence of a minimum density for stable compact objects in nature. This result arises rigorously from the generalized Buchdahl inequalities \cite{Boehmer:2005sm} and, following a similar analysis, the minimum charge-squared to mass ratio for electrically charged particles can also be obtained. Identifying this with the alternative expression for $r_{\rm e}$, obtained from the SNH, {\it also} yields (\ref{N-6}).

The third method, proposed by Beck \cite{Beck:2008rd} (2008) follows an axiomatic approach, based on analogy with the Kinchin axioms of information theory \cite{Khinchin}. These uniquely fix the form of the Shannon entropy, which forms the mathematical basis of statistical mechanics and thermodynamics, {\it without} making any assumptions about the underlying microphysical basis of these theories. Likewise, Beck's approach aims to uniquely fix the form of $\Lambda$ without reference to an underlying microphysical model. By formally replacing the dependence of the Shannon information measure $I$ on the probabilities of events $p_i$ by the dependence of $\Lambda$ on the fundamental constants of nature (i.e., the fundamental coupling constants, masses and mixing angles), Eq. (\ref{N-6}) was obtained from the requirements of `Fundamentality', `Boundedness', `Simplicity' and `Scale-invariance' \cite{Beck:2008rd}. 

A fourth derivation of Eq. (\ref{N-6}) has been proposed in a recent series of papers by Burikham, Cheamsawat, Harko, and Lake \cite{Burikham:2015sro,Lake:2017uzd,Burikham:2016cwz,Burikham:2017bkn} (2016-2017). This is based on the construction of a dark energy-modified minimum length uncertainty relation, dubbed the dark energy uncertainty principle, or DE-UP for short, which may be combined with {\it classical} minimum mass, radius and/or density bounds to yield stability conditions for self-gravitating, compact and quantum mechanical objects. Applying the DE-UP to charge-neutral particles recovers the minimum-mass bound previously obtained in \cite{Burikham:2015nma}, which is consistent with current bounds on the mass of the electron neutrino \cite{PlanckCollaboration}, whereas applying it to charged particles with $Q = \pm e$ yields the electron mass, in accordance with Eq. (\ref{N-6}). 

The form of the DE-UP proposed in \cite{Burikham:2015sro} leads naturally to holographic relation between the bulk and the de Sitter horizon, in which the number of minimum-volume `cells' equals  the number of Planck-sized `bits' on the boundary \cite{MenaMarugan:2001qn,Funkhouser:2005hp,Burikham:2016rbj}. However, it must be noted that, in order to maintain this relation for $\tau \ll \tau_0$, where $\tau_0 \simeq t_{\rm dS} = l_{\rm dS}/c$, we must substitute $l_{\rm dS} \rightarrow r_{\rm H}(\tau)$, where $r_{\rm H}(\tau)$ is the physical horizon radius at cosmic time $\tau$. Performing the same substitution in the DE-UP -- the validity of which is also supported by the physical arguments proposed in \cite{Lake:2017uzd} -- therefore gives rise to time-variation of the minimum mass for charge-neutral particles {\it and} of the maximum charge-squared to mass ratio for charged particles.  

At present, further analysis is required to determine whether the predictions of the DE-UP, which gives rise to a natural implementation of the LNH at the present epoch, as well as to a natural description of late-time accelerated expansion in terms of dark energy particles (see Sec. \ref{sect3.2}), is compatible with existing cosmological data. However, it must be noted that the present model implicitly assumes that $G$, $c$, $\hbar$ and $\Lambda$ are genuine universal constants which do not vary in time. Running of the gravitational coupling, as recently claimed in \cite{Gomez-Valent:2016nzf,Fritzsch:2016ewd,Sola:2016jky,Sola:2016vis}, or additional time-dependence in any of the parameters $\left\{G,c,\hbar,\Lambda\right\}$, may dramatically alter these predictions. With this in mind, we note that a model in which $\Lambda \propto \alpha_{\rm e}^{-6}$, in the context of varying $\alpha_{\rm e}$ cosmology, was considered in \cite{Wei:2016moy,Wei:2017mzf}, whereas alternative ways of incorporating the effects of Universal expansion/dark energy on the uncertainty principle were considered in \cite{Bambi:2008ts,Perivolaropoulos:2017rgq}.

Finally, we may consider the implications of a connection between ``dark" and ``light" physics, suggested by Eq. (\ref{N-6}), for the physics of black holes. {\it A priori}, Eqs. (\ref{Lambda})-(\ref{r_e}) and (\ref{N-6}) say nothing about black holes, yet if, as claimed in \cite{Burikham:2015sro,Lake:2017uzd}, the ultimate origin of these relations is the DE-UP, it is by no means clear whether this even applies to objects with masses $m \gtrsim m_{\rm Pl}$. In general, the form of positional uncertainty (if any) obeyed by the centre-of-mass of a black hole, remains an outstanding problem in contemporary theoretical physics. (See for \cite{CaMoPr:2011,CaMuNic:2014,Ca:2014,Lake:2015pma,Lake:2016did,Lake:2016enn} for recent works in this direction.) 

%Nonetheless, it is certainly worthwhile to attempt to extend the DE-UP into this region, which may be done na{\" i}vely by simply replacing the rest mass $m$ with the ADM mass $m_{\rm ADM} \simeq (m + m_{\rm Pl}^2/m)$. 
Nonetheless, it is certainly worthwhile to attempt to extend the DE-UP into this region, which may be done na{\" i}vely by simply replacing the rest mass $m$ with the `dual' ADM mass $m \rightarrow m_{\rm ADM}' \equiv m_{\rm Pl}^2/m_{\rm ADM} \simeq m_{\rm Pl}^2/(m + m_{\rm Pl}^2/m)$. This gives rise to a unified Compton-Schwarzschild line connecting the black hole and particle regimes (see \cite{CaMoPr:2011,CaMuNic:2014,Ca:2014,Lake:2015pma,Lake:2016did,Lake:2016enn} and \cite{Singh:2017wrb,Singh:2017ipg}). Since the DE-UP naturally implements holography in the $m \lesssim m_{\rm Pl}$ regime, it may be hoped that the extended version maintains it for $m \gtrsim m_{\rm Pl}$, which may have profound implications for the black hole information loss paradox \cite{Preskill:1992tc,Giddings:1995gd,Hawking:2005kf}. In this context, reassessing Beck's information-theoretic approach, subject to holographic constraints, may prove particularly fruitful. 

%%%%%%%%%%%%%%%%%%%%%%%%%%%%%%%%%%%%%%%%%%%%%%%%%%%
%%%%%%%%%%%%%%%%%%%%%%%%%%%%%%%%%%%%%%%%%%%%%%%%%%%
\section*{Acknowledgments}

M.L. thanks the Institute for Fundamental Study (IF), Phitsanulok, Thailand, the Yukawa Institute for Theoretical Physics (YITP), Kyoto, Japan, and the Asia-Pacific Center for Theoretical Physics (APCTP), Pohang, Korea, for their generous support of the $6^{\rm th}$ IF+YITP International Symposium, held at Naresuan University in August 2016. He is supported by a Naresuan University Research Fund individual research grant. 

%Biblio%%%%%%%%%%%%%%%%%%%%%%%%%%%%%%%%%%%%%%%%%%%%%%%%
%%%%%%%%%%%%%%%%%%%%%%%%%%%%%%%%%%%%%%%%%%%%%%%%%%%%


\begin{thebibliography}{99}

%DIRAC'S LNH (ORIGINAL PAPERS)

%\cite{Dirac1937}
\bibitem{Dirac1937}
  P.~A.~M.~Dirac,
  {\it The cosmological constants}, 
  Nature 139 (1937): 323.
  
%\cite{Dirac1938}
\bibitem{Dirac1938}
  P.~A.~M.~Dirac,  
  {\it A new basis for cosmology}, 
  Proc. R. Soc. Lond. A Vol. 165. No. 921 (1938).

%\cite{Dirac1974}
\bibitem{Dirac1974}
  P.~A.~M.~Dirac,
  {\it Cosmological Models and the Large Numbers Hypothesis}
  Proc. R. Soc. Lond. A 1974 338 439-446,
  DOI: 10.1098/rspa.1974.0095 (1974).

%\cite{Dirac1979}
\bibitem{Dirac1979}
  P.~A.~M.~Dirac,
  {\it The Large Numbers Hypothesis and the Einstein Theory of Gravitation}
  Proc. R. Soc. Lond. A 1979 365 19-30, 
  DOI: 10.1098/rspa.1979.0003 (1979).
  %\cite{Dirac1937,Dirac1974,Dirac1979} 
 
%DIRAC'S LNH (CONTEMPORARY REVIEW)
  
%\cite{Ray:2007cc}
\bibitem{Ray:2007cc} 
  S.~Ray, U.~Mukhopadhyay and P.~Pratim Ghosh,
  {\it Large Number Hypothesis: A Review},
  %Submitted to: Gen.Rel.Grav.
  [arXiv:0705.1836 [gr-qc]].
  
%COSMOLOGICAL CONCORDANCE MODEL  
  
%\cite{Ostriker:1995rn}
\bibitem{Ostriker:1995rn} 
  J.~P.~Ostriker and P.~J.~Steinhardt,
  {\it Cosmic concordance},
  astro-ph/9505066.
 
%ZELDOVICH LNH
  
%\cite{Zel'dovich:1968zz}
\bibitem{Zel'dovich:1968zz} 
  Y.~B.~Zel'dovich, A.~Krasinski and Y.~B.~Zeldovich,
  {\it The Cosmological constant and the theory of elementary particles},
  Sov.\ Phys.\ Usp.\  {\bf 11}, 381 (1968)
  [Gen.\ Rel.\ Grav.\  {\bf 40}, 1557 (2008)]
  [Usp.\ Fiz.\ Nauk {\bf 95}, 209 (1968)].
  doi:10.1007/s10714-008-0624-6, 10.1070/PU1968v011n03ABEH003927
 
%SDSS DATA
  
%\cite{Betoule:2014frx}
\bibitem{Betoule:2014frx} 
  M.~Betoule {\it et al.} [SDSS Collaboration],
  {\it Improved cosmological constraints from a joint analysis of the SDSS-II and SNLS supernova samples},
  Astron.\ Astrophys.\  {\bf 568}, A22 (2014)
  doi:10.1051/0004-6361/201423413
  [arXiv:1401.4064 [astro-ph.CO]].
 
%PLANCK DATA
 
%\cite{PlanckCollaboration}  
\bibitem{PlanckCollaboration} 
P.~A.~R.~Ade \textit{et al.} [Planck Collaboration],
{\it Planck 2013 results. I. Overview of products and scientific results},
arXiv:1303.5062 [astro-ph.CO]; 
P.~A.~R.~Ade \textit{et al.} [Planck Collaboration],
{\it Planck 2013 results. XVI. Cosmological parameters},
arXiv:1303.5076 [astro-ph.CO]. 
P.~A.~R.~Ade \textit{et al.} [Planck Collaboration], 
{\it Planck intermediate results. XXII. Frequency dependence of thermal emission from Galactic dust in intensity and polarization}
arXiv:1405.0874 [astro-ph.GA]; 
P.~A.~R.~Ade \textit{et al.} [Planck Collaboration], 
{\it Planck intermediate results. XXIX. All-sky dust modelling with Planck, IRAS, and WISE observations}
arXiv:1409.2495 [astro-ph.GA].
%\cite{Betoule:2014frx,PlanckCollaboration}

%THE HOLOGRAPHIC CONJECTURE  

%\cite{tHooft:1999rgb}
\bibitem{tHooft:1999rgb} 
  G.~'t Hooft,
  {\it The Holographic principle: Opening lecture},
  Subnucl.\ Ser.\  {\bf 37}, 72 (2001)
  doi:10.1142/9789812811585 0005
  [hep-th/0003004].
  
%\cite{Bousso:2002ju}
\bibitem{Bousso:2002ju} 
  R.~Bousso,
  {\it The Holographic principle},
  Rev.\ Mod.\ Phys.\  {\bf 74}, 825 (2002)
  doi:10.1103/RevModPhys.74.825
  [hep-th/0203101].
  %\cite{tHooft:1999rgb,Bousso:2002ju}

%NOTTALE

%\cite{Nottale}
\bibitem{Nottale}
L.~Nottale,
{\it Mach's Principle, Dirac's Large Number Hypothesis and the Cosmological Constant Problem} (preprint),
https://www.luth.obspm.fr/~luthier/nottale/arlambda.pdf (1993).

%BOEHMER AND HARKO (DE PARTICLES)
  
%\cite{Boehmer:2006fd}
\bibitem{Boehmer:2006fd} 
  C.~G.~Boehmer and T.~Harko,
  {\it Physics of dark energy particles},
  Found.\ Phys.\  {\bf 38}, 216 (2008)
  doi:10.1007/s10701-007-9199-4
  [gr-qc/0602081].

%WESSON (MASS QUANTIZATION)  

%\cite{Wesson:2003qn}
\bibitem{Wesson:2003qn}
  P.~S.~Wesson,
  {\it Is mass quantized?},
  Mod.\ Phys.\ Lett.\ A {\bf 19}, 1995 (2004).

%BOEHMER AND HARKO (COSMO. CONST. AND MIN. MASS)
  
%\cite{Boehmer:2005sm}
\bibitem{Boehmer:2005sm} 
  C.~G.~Boehmer and T.~Harko,
  {\it Does the cosmological constant imply the existence of a minimum mass?},
  Phys.\ Lett.\ B {\bf 630}, 73 (2005)
  doi:10.1016/j.physletb.2005.09.071
  [gr-qc/0509110].   

%MIN. MASS FOR NEUTRAL PARTICLES 
  
%\cite{Burikham:2015nma}
\bibitem{Burikham:2015nma}
  P.~Burikham, K.~Cheamsawat, T.~Harko and M.~J.~Lake,
  {\it The minimum mass of a spherically symmetric object in $D$-dimensions, and its implications for the mass hierarchy problem},
  Eur.\ Phys.\ J.\ C {\bf 75}, no. 9, 442 (2015)
  [arXiv:1508.03832 [gr-qc]].  
  
%EVIDENCE FOR GRANULARITY OF D.E. ON SCALES ~0.1 mm

%\cite{Perivolaropoulos:2016ucs}
\bibitem{Perivolaropoulos:2016ucs} 
  L.~Perivolaropoulos,
  {\it Sub-millimeter Spatial Oscillations of Newton's Constant: Theoretical Models and Laboratory Tests},
  arXiv:1611.07293 [gr-qc].

%GRAV. TURN-AROUND RADIUS
  
%\cite{Bhattacharya:2016vur}
\bibitem{Bhattacharya:2016vur} 
  S.~Bhattacharya, K.~F.~Dialektopoulos, A.~E.~Romano, C.~Skordis and T.~N.~Tomaras,
  {\it The maximum sizes of large scale structures in alternative theories of gravity},
  arXiv:1611.05055 [astro-ph.CO].

%MIN. MASS FOR CHARGED PARTICLES 
  
%\cite{Burikham:2015sro}
\bibitem{Burikham:2015sro}
  P.~Burikham, K.~Cheamsawat, T.~Harko and M.~J.~Lake,
  {\it The minimum mass of a charged spherically symmetric object in $D$ dimensions, its implications for fundamental particles, and holography},
  Eur.\ Phys.\ J.\ C {\bf 76}, no. 3, 106 (2016)
  [arXiv:1512.07413 [gr-qc]]. 

%MLUR FOR A DE UNIVERSE
 
%\cite{Lake:2017uzd}
\bibitem{Lake:2017uzd} 
  M.~J.~Lake,
  {\it Minimum length uncertainty relations for a dark energy Universe},
  arXiv:1712.00271 [gr-qc].
  %%CITATION = ARXIV:1712.00271;%%
    
%\cite{LakePaterek}
%\bibitem{LakePaterek}
%   M.~J.~Lake and T.~Paterek,
%   {\it Minimum length uncertainty relations for a dark energy Universe} (in preparation).  

%BECK'S AXIOMATIC APPROACH
   
%\cite{Beck:2008rd}
\bibitem{Beck:2008rd}
  C.~Beck,
  {\it Axiomatic approach to the cosmological constant},
  Physica A {\bf 388}, 3384 (2009).

%\cite{Khinchin}
\bibitem{Khinchin} 
  A. I. Khinchin, \textit{Mathematical Foundations of
  Information Theory}, Dover Publ., New York (1957).  
  
%APPLICATIONS TO MOD. GRAV. THEORIES  
   
%\cite{Burikham:2016cwz}
\bibitem{Burikham:2016cwz} 
  P.~Burikham, T.~Harko and M.~J.~Lake,
  {\it Mass bounds for compact spherically symmetric objects in generalized gravity theories},
  Phys.\ Rev.\ D {\bf 94}, no. 6, 064070 (2016)
  doi:10.1103/PhysRevD.94.064070
  [arXiv:1606.05515 [gr-qc]].
 
%\cite{Burikham:2017bkn}
\bibitem{Burikham:2017bkn} 
  P.~Burikham, T.~Harko and M.~J.~Lake,
  {\it The QCD mass gap and quark deconfinement scales as mass bounds in strong gravity},
  arXiv:1705.11174 [hep-th].
  %\cite{Burikham:2015sro,LakePaterek,Burikham:2016cwz,Burikham:2017bkn}
  
%REVIEWS OF MLUR IN Q. GRAV.
  
%\cite{Hossenfelder:2012jw}
\bibitem{Hossenfelder:2012jw}  S.~Hossenfelder,
  {\it Minimal Length Scale Scenarios for Quantum Gravity},
  Living Rev.\ Rel.\ \textbf{16}, 2 (2013).
  
%\cite{Garay:1994en}
\bibitem{Garay:1994en} 
  L.~J.~Garay,
  {\it Quantum gravity and minimum length},
  Int.\ J.\ Mod.\ Phys.\ A {\bf 10}, 145 (1995)
  %doi:10.1142/S0217751X95000085
  [gr-qc/9403008].
  %\cite{Hossenfelder:2012jw,Garay:1994en}  
  
%MLUR IN CANONICAL QM (RIGOROUS DERIVATION FROM HEISENBERG PIC.)  

%\cite{Calmet:2004mp}
\bibitem{Calmet:2004mp}  X.~Calmet, M.~Graesser and S.~D.~H.~Hsu,
  {\it Minimum length from quantum mechanics and general relativity},
Phys.\ Rev.\ Lett.\ \textbf{93}, 211101 (2004).

%\cite{Calmet:2005mh}
\bibitem{Calmet:2005mh}  X.~Calmet, M.~Graesser and S.~D.~H.~Hsu,
  {\it Minimum length from first principles},
  Int.\ J.\ Mod.\ Phys.\ D \textbf{14}, 2195 (2005).
  %\cite{Calmet:2004mp,Calmet:2005mh}
  
%SALECKER & WIGNER GEDANKEN EXPT.

%\cite{Salecker:1957be}
\bibitem{Salecker:1957be} 
  H.~Salecker and E.~P.~Wigner,
  {\it Quantum limitations of the measurement of space-time distances},
  Phys.\ Rev.\  {\bf 109}, 571 (1958).
  doi:10.1103/PhysRev.109.571  
  
%BEKENSTEIN CHARGED PARTICLE BOUND AND ITS EXTENSIONS
 
%\cite{Bekenstein:1971ej}
\bibitem{Bekenstein:1971ej}
  J.~D.~Bekenstein,
  {\it Hydrostatic Equilibrium and Gravitational Collapse of Relativistic Charged Fluid Balls},
  Phys.\ Rev.\ D {\bf 4}, 2185 (1971). 
  
%\cite{Boehmer:2007gq}
\bibitem{Boehmer:2007gq}
  C.~G.~Boehmer and T.~Harko,
  {\it Minimum mass-radius ratio for charged gravitational objects},
  Gen.\ Rel.\ Grav.\  {\bf 39}, 757 (2007).
   %\cite{Bekenstein:1971ej,Boehmer:2007gq} 

%STRONG GRAVITY PAPERS
   
%\cite{Isham:1971gm}
\bibitem{Isham:1971gm} 
  C.~J.~Isham, A.~Salam and J.~A.~Strathdee,
  {\it F-dominance of gravity},
  Phys.\ Rev.\ D {\bf 3}, 867 (1971).
  doi:10.1103/PhysRevD.3.867

 %\cite{Isham:1970aw}
\bibitem{Isham:1970aw} 
  C.~J.~Isham, A.~Salam and J.~A.~Strathdee,
  {\it Infinity suppression gravity modified quantum electrodynamics},
  Phys.\ Rev.\ D {\bf 3}, 1805 (1971).
  doi:10.1103/PhysRevD.3.1805
 
%\cite{Isham:1973jh}
\bibitem{Isham:1973jh} 
  C.~J.~Isham, A.~Salam and J.~A.~Strathdee,
  {\it 2+ nonet as gauge particles for sl(6,c) symmetry},
  Phys.\ Rev.\ D {\bf 8}, 2600 (1973).
  doi:10.1103/PhysRevD.8.2600
  
%\cite{Isham:1974wx}
\bibitem{Isham:1974wx} 
  C.~J.~Isham and M.~C.~Tucker,
  {\it A $U(3)$ Version of the f-g Theory},
  Phys.\ Rev.\ D {\bf 10}, 3219 (1974).
  doi:10.1103/PhysRevD.10.3219  
  %\cite{Isham:1971gm,Isham:1970aw,Isham:1973jh,Isham:1974wx}

%NG's MLUR FOR AN EXPANDING UNIVERSE

%\cite{Ng:2016wzj}
\bibitem{Ng:2016wzj} 
  Y.~J.~Ng,
  {\it Quantum Foam, Gravitational Thermodynamics, and the Dark Sector},
  J.\ Phys.\ Conf.\ Ser.\  {\bf 845}, no. 1, 012001 (2017)
  doi:10.1088/1742-6596/845/1/012001
  [arXiv:1701.00017 [gr-qc]].

%RUNNING GRAVITATIONAL CONSTANT MODELS

%\cite{Gomez-Valent:2016nzf}
\bibitem{Gomez-Valent:2016nzf}
  A.~G{\'o}mez-Valent, J.~Sol{\`a} and J.~d.~C.~P{\'e}rez,
  {\it Running vacuum versus the $\Lambda$CDM},
  arXiv:1605.06448 [gr-qc].

%\cite{Fritzsch:2016ewd}
\bibitem{Fritzsch:2016ewd}
  H.~Fritzsch, R.~C.~Nunes and J.~Sol{\`a},
  {\it Running vacuum in the Universe and the time variation of the fundamental constants of Nature},
  arXiv:1605.06104 [hep-ph].

%\cite{Sola:2016jky}
\bibitem{Sola:2016jky}
  J.~Sol{\`a}, A.~G{\'o}mez-Valent and J.~d.~C.~P{\'e}rez,
  {\it First evidence of running cosmic vacuum: challenging the concordance model},
  arXiv:1602.02103 [astro-ph.CO].

%\cite{Sola:2016vis}
\bibitem{Sola:2016vis}
  J.~Sol{\`a},
  {\it Running Vacuum in the Universe: current phenomenological status},
  arXiv:1601.01668 [gr-qc].
  %\cite{Gomez-Valent:2016nzf,Fritzsch:2016ewd,Sola:2016jky,Sola:2016vis}

%VSL 

%\cite{Albrecht:1998ir}
\bibitem{Albrecht:1998ir} 
  A.~Albrecht and J.~Magueijo,
  {\it A Time varying speed of light as a solution to cosmological puzzles},
  Phys.\ Rev.\ D {\bf 59}, 043516 (1999)
  doi:10.1103/PhysRevD.59.043516
  [astro-ph/9811018].

%\cite{Barrow:1998eh}
\bibitem{Barrow:1998eh} 
  J.~D.~Barrow,
  {\it Cosmologies with varying light speed},
  astro-ph/9811022.
  
%\cite{Albrecht:1998ij}
\bibitem{Albrecht:1998ij} 
  A.~Albrecht,
  {\it Cosmology with a time varying speed of light},
  AIP Conf.\ Proc.\  {\bf 478}, 263 (1999)
  doi:10.1063/1.59450
  [astro-ph/9904185].
  
%\cite{Moffat:2002nm}
\bibitem{Moffat:2002nm} 
  J.~W.~Moffat,
  {\it Variable speed of light cosmology: An Alternative to inflation},
  hep-th/0208122.

%\cite{Qi:2014zja}
\bibitem{Qi:2014zja} 
  J.~Z.~Qi, M.~J.~Zhang and W.~B.~Liu,
  {\it Observational constraint on the varying speed of light theory},
  Phys.\ Rev.\ D {\bf 90}, no. 6, 063526 (2014)
  doi:10.1103/PhysRevD.90.063526
  [arXiv:1407.1265 [gr-qc]].
  
%\cite{Landau:2000mx}
\bibitem{Landau:2000mx} 
  S.~Landau, P.~D.~Sisterna and H.~Vucetich,
  {\it Charge conservation and time varying speed of light},
  Phys.\ Rev.\ D {\bf 63}, 081303 (2001)
  doi:10.1103/PhysRevD.63.081303
  [astro-ph/0007108].
  
%\cite{Magueijo:2003gj}
\bibitem{Magueijo:2003gj} 
  J.~Magueijo,
  {\it New varying speed of light theories},
  Rept.\ Prog.\ Phys.\  {\bf 66}, 2025 (2003)
  doi:10.1088/0034-4885/66/11/R04
  [astro-ph/0305457].
  
%\cite{Dabrowski:2016lot}
\bibitem{Dabrowski:2016lot} 
  M.~P.~Dabrowski, V.~Salzano, A.~Balcerzak and R.~Lazkoz,
  {\it New tests of variability of the speed of light},
  EPJ Web Conf.\  {\bf 126}, 04012 (2016).
  doi:10.1051/epjconf/201612604012
  %\cite{Albrecht:1998ir,Barrow:1998eh,Albrecht:1998ij,Moffat:2002nm,Qi:2014zja,Landau:2000mx,Magueijo:2003gj,Dabrowski:2016lot}

 %DARK ENERGY (REVIEWS)

%\cite{Tsujikawa:2010zz}
\bibitem{Tsujikawa:2010zz} 
  S.~Tsujikawa,
  {\it Recent status of dark energy},
  Mod.\ Phys.\ Lett.\ A {\bf 25}, 843 (2010).
  doi:10.1142/S0217732310000010

%\cite{AmendolaTsujikawa(2010)}
 \bibitem{AmendolaTsujikawa(2010)}
  L.~Amendola and S.~Tsujikawa, 
  {\it Dark Energy: Theory and Observations},
  Cambridge University Press, 2010.~ISBN: 9780521516006.
  
%\cite{Li:2013hia}
\bibitem{Li:2013hia} 
  M.~Li, X.~D.~Li, S.~Wang and Y.~Wang,
  {\it Dark Energy},
  The Universe {\bf 1}, no. 4, 24 (2013).

%\cite{Li:2012dt}
\bibitem{Li:2012dt} 
  M.~Li, X.~D.~Li, S.~Wang and Y.~Wang,
  {\it Dark Energy: A Brief Review},
  Front.\ Phys.\ (Beijing) {\bf 8}, 828 (2013)
  doi:10.1007/s11467-013-0300-5
  [arXiv:1209.0922 [astro-ph.CO]].
%\cite{Tsujikawa:2010zz,AmendolaTsujikawa(2010),Li:2013hia,Li:2012dt}
  
%VARYING ALPHA MODELS

%\cite{Barrow:1998df}
\bibitem{Barrow:1998df} 
  J.~D.~Barrow and J.~Magueijo,
  {\it Varying alpha theories and solutions to the cosmological problems},
  Phys.\ Lett.\ B {\bf 443}, 104 (1998)
  doi:10.1016/S0370-2693(98)01294-5
  [astro-ph/9811072].
  
%\cite{Barrow:1999st}
\bibitem{Barrow:1999st} 
  J.~D.~Barrow and J.~Magueijo,
  {\it Can a changing alpha explain the supernovae results?},
  Astrophys.\ J.\  {\bf 532}, L87 (2000)
  doi:10.1086/312572
  [astro-ph/9907354].
  
%\cite{Moffat:2001sf}
\bibitem{Moffat:2001sf} 
  J.~W.~Moffat,
  {\it A Model of varying fine structure constant and varying speed of light},
  astro-ph/0109350.
  
%\cite{Kozlov:2008iy}
\bibitem{Kozlov:2008iy} 
  M.~G.~Kozlov, S.~G.~Porsev, S.~A.~Levshakov, D.~Reimers and P.~Molaro,
  {\it Mid- and far-infrared fine-structure line sensitivities to hypothetical variability of the fine-structure constant},
  Phys.\ Rev.\ A {\bf 77}, 032119 (2008)
  doi:10.1103/PhysRevA.77.032119
  [arXiv:0802.0269 [astro-ph]].
  
 %\cite{Barrow:2014vva}
\bibitem{Barrow:2014vva} 
  J.~D.~Barrow and J.~Magueijo,
  {\it Local Varying-Alpha Theories},
  Mod.\ Phys.\ Lett.\ A {\bf 30}, no. 22, 1540029 (2015)
  doi:10.1142/S0217732315400295
  [arXiv:1412.3278 [gr-qc]].
  %\cite{Barrow:1998df,Barrow:1999st,Moffat:2001sf,Kozlov:2008iy,Barrow:2014vva}
  
%EFFECT OF VARYING ALPHA ON COSMIC STRING PHENOM.  
  
%\cite{Lake:2010wt}
\bibitem{Lake:2010wt} 
  M.~Lake and J.~Ward,
  {\it A Generalisation of the Nielsen-Olesen Vortex: Non-cylindrical strings in a modified Abelian-Higgs model},
  JHEP {\bf 1104}, 048 (2011)
  doi:10.1007/JHEP04(2011)048
  [arXiv:1009.2104 [hep-ph]].
  
%\cite{Lake:2010qsa}
\bibitem{Lake:2010qsa} 
  M.~J.~Lake,
  {\it Cosmic necklaces in string theory and field theory},
  Ph.D. Thesis, Queen Mary, University of London (2010).
  %\cite{Lake:2010wt,Lake:2010qsa}

%GENERAL VARYING CONSTANT THEORIES

%\cite{Brandenberger:1999bi} 
\bibitem{Brandenberger:1999bi} 
  R.~H.~Brandenberger and J.~Magueijo,
  {\it Imaginative cosmology},
  hep-ph/9912247.
  
%\cite{Gopakumar:2000kp}
\bibitem{Gopakumar:2000kp} 
  P.~Gopakumar and G.~V.~Vijayagovindan,
  {\it Solutions to cosmological problems with energy conservation and varying c, G and $\Lambda$},
  Mod.\ Phys.\ Lett.\ A {\bf 16}, 957 (2001)
  doi:10.1142/S0217732301004042
  [gr-qc/0003098].
  
 %\cite{Biesiada:2003tp}
\bibitem{Biesiada:2003tp} 
  M.~Biesiada,
  {\it Varying fundamental constants: A dynamical systems approach},
  Astrophys.\ Space Sci.\  {\bf 283}, 511 (2003).
  doi:10.1023/A:1022564903718
  
%\cite{Barrow:2005sv}
\bibitem{Barrow:2005sv} 
  J.~D.~Barrow,
  {\it Cosmological bounds on spatial variations of physical constants},
  Phys.\ Rev.\ D {\bf 71}, 083520 (2005)
  doi:10.1103/PhysRevD.71.083520
  [astro-ph/0503434].
  
 %\cite{Scoccola:2009zh}
\bibitem{Scoccola:2009zh} 
  C.~G.~Scoccola,
  {\it Variation of the fine structure constant and the electron mass at early Universe},
  arXiv:0906.0329 [astro-ph.CO].
  %\cite{Brandenberger:1999bi,Gopakumar:2000kp,iesiada:2003tp,Barrow:2005sv,Scoccola:2009zh} 
  
%MODELS OF DARK ENERGY COUPLING TO EM

%\cite{Yan:2010nx}
\bibitem{Yan:2010nx} 
  M.~L.~Yan,
  {\it One Electron Atom in Special Relativity with de Sitter Space-Time Symmetry},
  Commun.\ Theor.\ Phys.\  {\bf 57}, 930 (2012)
  doi:10.1088/0253-6102/57/6/04
  [arXiv:1004.3023 [physics.gen-ph]].
  
%\cite{Narimani:2011rb}
\bibitem{Narimani:2011rb} 
  A.~Narimani, A.~Moss and D.~Scott,
  {\it Dimensionless cosmology},
  Astrophys.\ Space Sci.\  {\bf 341}, 617 (2012)
  doi:10.1007/s10509-012-1113-7
  [arXiv:1109.0492 [astro-ph.CO]].
  
%\cite{Hollenstein:2012mb}
\bibitem{Hollenstein:2012mb} 
  L.~Hollenstein, R.~K.~Jain and F.~R.~Urban,
  {\it Cosmological Ohm's law and dynamics of non-minimal electromagnetism},
  JCAP {\bf 1301}, 013 (2013)
  doi:10.1088/1475-7516/2013/01/013
  [arXiv:1208.6547 [astro-ph.CO]].
  
%\cite{Pandolfi:2014jka}
\bibitem{Pandolfi:2014jka} 
  S.~Pandolfi,
  {\it Dark Energy coupling with electromagnetism as seen from future low-medium redshift probes},
  J.\ Phys.\ Conf.\ Ser.\  {\bf 566}, no. 1, 012005 (2014).
  doi:10.1088/1742-6596/566/1/012005  
    
%\cite{Martins:2015jta}
\bibitem{Martins:2015jta} 
  C.~J.~A.~P.~Martins, A.~M.~M.~Pinho, R.~F.~C.~Alves, M.~Pino, C.~I.~S.~A.~Rocha and M.~von Wietersheim,
  {\it Dark energy and Equivalence Principle constraints from astrophysical tests of the stability of the fine-structure constant},
  JCAP {\bf 1508}, no. 08, 047 (2015)
  doi:10.1088/1475-7516/2015/08/047
  [arXiv:1508.06157 [astro-ph.CO]].
  %\cite{Yan:2010nx,Narimani:2011rb,Hollenstein:2012mb,Pandolfi:2014jka,Martins:2015jta}  

%LAMBDA ~ ALPHA^{-6} IN VARYING ALPHA COSMOLOGY  

%\cite{Wei:2016moy}
\bibitem{Wei:2016moy} 
  H.~Wei, X.~B.~Zou, H.~Y.~Li and D.~Z.~Xue,
  {\it Cosmological Constant, Fine Structure Constant and Beyond},
  Eur.\ Phys.\ J.\ C {\bf 77}, no. 1, 14 (2017)
  doi:10.1140/epjc/s10052-016-4581-z
  [arXiv:1605.04571 [gr-qc]].
  
%\cite{Wei:2017mzf}
\bibitem{Wei:2017mzf} 
  H.~Wei and D.~Z.~Xue,
  {\it Observational Constraints on Varying Alpha in $\Lambda(\alpha)$CDM Cosmology},'
  arXiv:1706.04063 [gr-qc].
  %\cite{Wei:2016moy,Wei:2017mzf}

%HOLOGRAPHY AND THE LNH
 
%\cite{MenaMarugan:2001qn}
\bibitem{MenaMarugan:2001qn} 
  G.~A.~Mena Marugan and S.~Carneiro,
  {\it Holography and the large number hypothesis},
  Phys.\ Rev.\ D {\bf 65}, 087303 (2002)
  doi:10.1103/PhysRevD.65.087303
  [gr-qc/0111034].
    
%\cite{Funkhouser:2005hp}
\bibitem{Funkhouser:2005hp}
  S.~Funkhouser,
  {\it The Large number coincidence, the cosmic coincidence and the critical acceleration},
  Proc.\ Roy.\ Soc.\ Lond.\ A {\bf 462}, 3657 (2006).
   
%\cite{Burikham:2016rbj}
\bibitem{Burikham:2016rbj}
  P.~Burikham, R.~Dhanawittayapol and T.~Wuthicharn,
  {\it A new mass scale, implications on black hole evaporation and holography},
  Int.\ J.\ Mod.\ Phys.\  {\bf 31}, no. 16, 1650089 (2016).
  %\cite{MenaMarugan:2001qn,Funkhouser:2005hp,Burikham:2016rbj} 
   
%BAMBI'S ALT. REVISION OF THE GUP (INC. DE SITTER RADIUS)
  
%\cite{Bambi:2008ts}
\bibitem{Bambi:2008ts} 
  C.~Bambi,
  {\it A Revision of the Generalized Uncertainty Principle},
  Class.\ Quant.\ Grav.\  {\bf 25}, 105003 (2008)
  doi:10.1088/0264-9381/25/10/105003
  [arXiv:0804.4746 [gr-qc]].
  
%PERIVAROLOPOULOS' MLUR FOR COSMO. HORIZONS

%\cite{Perivolaropoulos:2017rgq}
\bibitem{Perivolaropoulos:2017rgq} 
  L.~Perivolaropoulos,
  {\it Cosmological Horizons, Uncertainty Principle and Maximum Length Quantum Mechanics},
  arXiv:1704.05681 [gr-qc].
  
%DIFFERENT POSITIONAL UNCERTAINY FOR BH?
  
%\cite{CaMoPr:2011}
\bibitem{CaMoPr:2011}
  B.~Carr, L.~Modesto and I.~Premont-Schwarz,  
  {\it Generalized Uncertainty Principle and Self-dual Black Holes},
  arXiv: 1107.0708 [gr-qc].

%\cite{CaMuNic:2014}
\bibitem{CaMuNic:2014}
  B.~Carr,  J. Mureika and P. Nicolini,
  {\it Sub-Planckian black holes and the Generalized Uncertainty Principle}, 
  JHEP 07, 052 (2016), arXiv:1504.07637.

%\cite{Ca:2014}
\bibitem{Ca:2014}
  B.~Carr,  
  {\it The Black Hole Uncertainty Principle Correspondence}, 
  In 1st Karl Schwarzschild Meeting on Gravitational Physics, eds. P. Nicolini, M. Kaminski, J. Mureika, M. Bleicher, pp 159-167 (Springer), arXiv:1402:1427 (2015).

%\cite{Lake:2015pma}
\bibitem{Lake:2015pma} 
  M.~J.~Lake and B.~Carr,
  {\it The Compton-Schwarzschild correspondence from extended de Broglie relations},
  JHEP 1511, 105 (2015),
  doi:10.1007/JHEP11(2015)105
  arXiv:1505.06994 [gr-qc].
  
%\cite{Lake:2016did}
\bibitem{Lake:2016did} 
  M.~J.~Lake,
  {\it Which quantum theory must be reconciled with gravity? (And what does it mean for black holes?)},
  Universe {\bf 2}, no. 4, 24 (2016)
  doi:10.3390/universe2040024
  [arXiv:1607.03689 [gr-qc]].
  
%\cite{Lake:2016enn}
\bibitem{Lake:2016enn} 
  M.~J.~Lake and B.~Carr,
  {\it The Compton-Schwarzschild relations in higher dimensions},
  arXiv:1611.01913 [gr-qc].
  %\cite{CaMoPr:2011,CaMuNic:2014,Ca:2014,Lake:2015pma,Lake:2016did,Lake:2016enn}
  
%\cite{Singh:2017wrb}
\bibitem{Singh:2017wrb} 
  T.~P.~Singh,
  {\it A new length scale for quantum gravity},
  arXiv:1704.00747 [gr-qc].
  
%\cite{Singh:2017ipg}
\bibitem{Singh:2017ipg} 
  T.~P.~Singh,
  {\it A new length scale, and modified Einstein-Cartan-Dirac equations for a point mass},
  arXiv:1705.05330 [gr-qc].
  %\cite{Singh:2017wrb,Singh:2017ipg}
  
%BLACK HOLE INFO. LOSS PARADOX

%\cite{Preskill:1992tc}
\bibitem{Preskill:1992tc} 
  J.~Preskill,
  {\it Do black holes destroy information?},'
  In *Houston 1992, Proceedings, Black holes, membranes, wormholes and superstrings* 22-39, and Caltech Pasadena - CALT-68-1819 (92,rec.Oct.) 17 p
  [hep-th/9209058].

%\cite{Giddings:1995gd}
\bibitem{Giddings:1995gd} 
  S.~B.~Giddings,
  {\it The Black hole information paradox},
  hep-th/9508151.

%\cite{Hawking:2005kf}
\bibitem{Hawking:2005kf} 
  S.~W.~Hawking,
  {\it Information loss in black holes},
  Phys.\ Rev.\ D {\bf 72}, 084013 (2005)
  doi:10.1103/PhysRevD.72.084013
  [hep-th/0507171].
  %\cite{Preskill:1992tc,Giddings:1995gd,Hawking:2005kf}
  
%DIRAC'S EXTENSIBLE MODEL OF THE ELECTRON

%\cite{Dirac:1962iy}
%\bibitem{Dirac:1962iy} 
%  P.~A.~M.~Dirac,
%  {\it An Extensible model of the electron},
%  Proc.\ Roy.\ Soc.\ Lond.\ A {\bf 268}, 57 (1962).
%  doi:10.1098/rspa.1962.0124
  
\end{thebibliography}
\end{document}